\documentclass[a4paper,english]{iopart}
\usepackage[T1]{fontenc}
\usepackage[latin9]{inputenc}
\usepackage{graphicx}

\usepackage{iopams}
\usepackage{setstack}

\usepackage{babel}

\begin{document}

\title{Post-Tanner stages of droplet spreading: the energy balance approach
revisited }

\author{S Mechkov$^{1}$, A-M Cazabat$^{2}$ and G Oshanin$^{1,3}$}

\address{$^{1}$Laboratoire de Physique Théorique de la Matière Condensée,
Université Pierre et Marie Curie, 4 place Jussieu, 75252 Paris Cedex
5 France}

\address{$^{2}$Laboratoire de Physique Statistique, Ecole Normale Supérieure,
75252 Paris Cedex 5 France}

\address{$^{3}$Laboratory J.-V. Poncelet (UMI CNRS 2615), Independent University
of Moscow, Bolshoy Vlasyevskiy Pereulok 11, 119002 Moscow, Russia
}

\ead{mechkov@lptmc.jussieu.fr, anne-marie.cazabat@lps.ens.fr,  oshanin@lptmc.jussieu.fr}
\begin{abstract}
The spreading of a circular liquid drop on a solid substrate can be
described by the time evolution of its base radius $R(t)$. In complete
wetting the quasistationary regime (far away from initial and final
transients) typically obeys the so-called Tanner law, with $R\sim t^{\alpha_{\mathrm{T}}}$,
$\alpha_{\mathrm{T}}=1/10$. Late-time spreading may differ significantly
from the Tanner law: in some cases the drop does not thin down to
a molecular film and instead reaches an equilibrium pancake-like shape;
in other situations, as revealed by recent experiments with spontaneously
spreading nematic crystals, the growth of the base radius accelerates
after the Tanner stage. Here we demonstrate that these two seemingly
conflicting trends can be reconciled within a suitably revisited energy
balance approach, by taking into account the line tension contribution
to the driving force of spreading: a positive line tension is responsible
for the formation of pancake-like structures, whereas a negative line
tension tends to lengthen the contact line and induces an accelerated
spreading (a transition to a faster power law for $R(t)$ than in
the Tanner stage).
\end{abstract}

\vspace{2pc}

\noindent \vspace{2pc}

\submitto{\JPCM}

\maketitle

\section{Introduction}

The spreading of a liquid on a solid surface is a complicated process
where many factors come into play, not necessarily known and not always
controllable: the kinetic behavior may be strongly influenced or even
dominated by the volatility and viscosity of the liquid (or by other
rheological parameters if the liquid is non-Newtonian), by the presence
of impurities in the bulk phases (chemical contaminants, surfactants,
polymers, etc), by the roughness or texture of the surface, or by
its crystalline structure and chemical composition. In consequence,
the spreading of thin films is generally dependent upon the details
of the structures and interactions in the co-existing phases. By contrast,
for thicker films and drops one expects a less specific behavior,
described by universal laws. However, although the prominent features 
of the spreading of macroscopic drops are relatively well understood 
\cite{degennes85,cazabat,joanny}, a comprehensive theoretical framework
in which all of the experimentally observed phenomena harmoniously
find their place is still lacking at present.

Here we are concerned with a standard textbook problem - the spontaneous
spreading of a non-volatile drop on an ideal, flat, clean, horizontal,
homogeneous solid surface (see figure 1). The drop has a macroscopic
size but is sufficiently small that we can completely discard the
effects of gravity. The spreading parameter $S=\sigma_{\mathrm{SG}}-\sigma_{\mathrm{SL}}-\sigma$
is positive, so that the drop tends to cover as much of the solid
surface as possible to shield it against the gas phase; $\sigma_{\mathrm{SG}}$,
$\sigma_{\mathrm{SL}}$ and $\sigma$ are the interfacial tensions
of the solid/gas, solid/liquid and liquid/gas interfaces, respectively.

\vspace{0.1in}

We focus on the following three known features of spontaneous spreading:
\begin{itemize}
\item The base radius $R(t)$ of a circular drop grows during spreading
and, at intermediate times $t$ (far from initial and final transients),
typically obeys a power law $R(t)\sim t^{\alpha_{T}}$ with $\alpha_{\mathrm{T}}=1/10$
\cite{voinov,tanner}, known as the Tanner law.
\item The final stage of spreading of non-volatile droplets is not always
a molecular film. Sometimes a flat, bounded structure is reached instead
- a so-called pancake (see \cite{degennes85} and \cite{ruck,joanny1984,joanny1984-1}).
\item An acceleration of the spreading process (an apparent transition from
Tanner's power law to a faster one) has been observed for spontaneously
spreading nematic liquid crystals \cite{cazb,cazc}. Experiments revealed
an algebraic growth $R(t)\sim t^{\alpha}$ with $\alpha$ nearly twice
as large as the exponent $\alpha_{\mathrm{T}}$ characterizing the
Tanner law: $\alpha=0.2$ \cite{cazb} and $\alpha=0.19$ \cite{cazc}.
\end{itemize}
At present time, it is well understood why the radius $R(t)$ of a 
spontaneously spreading circular drop grows in proportion to $t^{1/10}$. 
This law has been derived analytically \cite{voinov,tanner,hervet} and 
confirmed experimentally on many accounts \cite{tanner,aus,caz,we}.
The fundamental
argument is that the hydrodynamics in the bulk of a drop are driven
by capillary forces alone, which directly yields $R\sim t^{1/10}$
for a self-similar bulk, in the lubrication approximation \cite{voinov,tanner}.
Alternatively, the trend can be regarded as a competition between
the hydrodynamic dissipation (primarily in the contact line region
of the drop) and an unbalanced capillary force \cite{degennes85,cazabat,joanny,hervet}.
Note that the law is rather universal - in the sense that the observed
exponent $1/10$ is most often independent of the precise nature of the 
spreading liquid - and has been observed not only for simple liquids, but 
for oils, polymeric liquids, liquid metals and nematic liquid crystals. 
For non-Newtonian liquids some deviations from the Tanner law can be observed, 
attributable to their specific rheological properties and hence, specific 
features of the hydrodynamic dissipation in the bulk.

The reason why in some cases a spontaneously spreading droplet attains
an equilibrium pancake-like form is also clear. Such flat pancake-like
structures are sometimes more favorable energetically than molecular
films: this occurs when short-range interactions promote dewetting,
even though the overall situation is that of complete wetting. Theoretically,
such structures have been predicted and analyzed in \cite{degennes85}
and \cite{ruck,joanny1984}. They were also observed experimentally
(see, e.g., \cite{pancakes}). A key feature of the spreading process
as it reaches a pancake shape - as a transient following the Tanner
stage - is that the base radius of the drop tends to a stationary
value (see figure 1b). It must be noted, though, that the shape of 
such a drop differs considerably from that of a capillary cap; 
the geometrical meaning of $R(t)$ is also quite different.

\begin{figure}
\begin{centering}
\includegraphics[width=0.8\columnwidth]{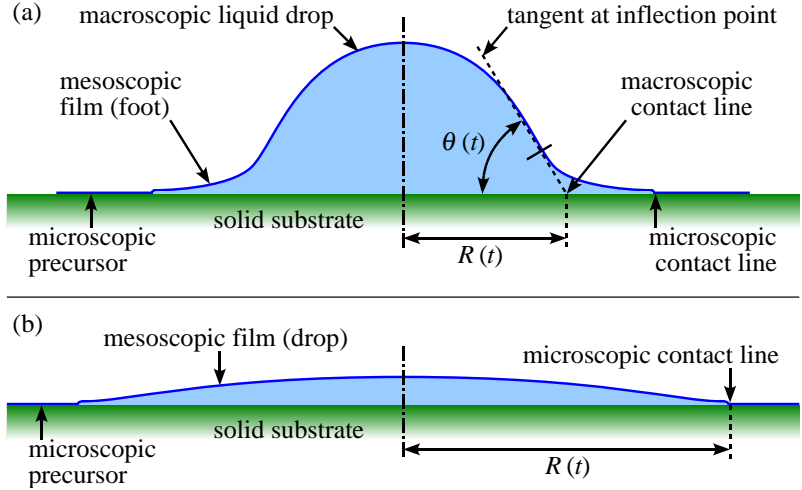}
\par\end{centering}

\caption{\label{fig:DropCrossSection} a) A sketch of a macroscopic liquid droplet spreading
on a solid substrate. $R(t)$ and $\theta(t)$ denote the base radius
and the contact angle of the macroscopic part of the droplet, respectively,
as inferred from the inflection point at the apparent contact line.
b) A sketch of a liquid droplet at a late spreading time, in the situation
where the final stage of spreading is a mesoscopic pancake. An inflexion
point may still exist, but the relevant $R(t)$ now corresponds to
the limit between the microscopic and mesoscopic regions; the corresponding
$\theta(t)$ is undefined.}

\end{figure}

As opposed to the Tanner law and the emergence of mesoscopic pancakes,
the physical origin of the accelerated spreading observed in \cite{cazb,cazc}
has yet to be clarified. Also, the latter trend is apparently in conflict
with the notion that a Tanner stage must be terminated by the onset
of either a molecular film or a mesoscopic pancake.

Our motivation in this paper is to identify the essential factors
of the spreading kinetics that might have been disregarded so far,
and thus to achieve a more complete qualitative understanding of the
standard textbook problem. We focus more specifically on the spreading
dynamics that may develop after the Tanner stage and try to account
for the two apparently conflicting trends, possibly by introducing
new notions or mechanisms. As a first step in this direction, we perform
an analysis within the classical framework of an energy balance approach
\cite{degennes85}. This approach is known to provide a qualitative
derivation of the Tanner law, capturing only a few essential features
of the phenomenon, in a physically transparent fashion. As noted in
\cite{berg}, the energy-based equations are functionally equivalent
to the standard hydrodynamic approaches used in the literature but
are lighter in terms of analytical calculations and assumptions involved (as
compared to, e.g., phenomenological boundary conditions, should the
thickness profile be described by a differential equation).

Within this approach, the Tanner law is obtained by balancing the
rate of energy dissipation in the spreading macroscopic droplet and the driving
force of spreading, which is taken equal to the unbalanced Young force.
We point out that the analysis of de Gennes \cite{degennes85} disregards
the line tension contribution to the driving force of spreading \cite{getta,schimmele07}.
Typically, the line tension $\tau$ is very small - only $10^{-10}$
to $10^{-11}$ N (see, e.g., \cite{herm,herm2,herm3}) - and it is
legitimate to neglect it when dealing with large, essentially capillary
droplets. Nonetheless we show that, when such a contribution is taken
into account, a consistent, non-conflicting picture emerges, with
the following trends for a drop (supposedly well-approximated by a
spherical cap):
\begin{itemize}
\item At sufficiently early spreading times the effect of $\tau$ is negligible
and the Tanner stage holds.
\item At long times and for negative values of $\tau$, the spreading process
crosses over to a significantly faster power law than Tanner's $R\sim t^{1/10}$
(as observed for the nematic droplets \cite{cazb,cazc}).
\item At long times and for positive values of $\tau$, the growth of $R(t)$
slows down and terminates at a finite value $R(\infty)$. This latter
trend is indicative of the emergence of pancakes.
\end{itemize}
Therefore, the approach presented in this paper resolves a seemingly
controversial behavior of spreading processes, provided that a properly
defined line tension $\tau$ is taken into account. We note, however, that this 
approach is justified only in the case of macroscopic drops, for which both the 
surface tension and the line tension are valid notions.

The paper is outlined as follows. In section 2 we first present the
derivation of the Tanner law in terms of the energy balance approach.
Then, in section 3, we revisit the standard picture by analyzing different
factors which may influence the spreading kinetics, especially the
notion of line tension. Finally, in section 4 we discuss possible
limitations of our approach.

\section{Energy balance approach}

To lay the basis of our analysis, we start with the derivation of
the Tanner law within the framework of the energy balance approach,
originally presented by de Gennes in his 1985 review paper \cite{degennes85}.
Additional details, discussions and applications of this approach
can be found in \cite{we,berg}. Further on, in section 3, we will
discuss a couple of additional factors which are missing from the
seminal approach, but may account for the abnormal spreading behavior
observed after the Tanner stage.

In figure 1a we sketched a typical configuration for a macroscopic liquid 
droplet spreading spontaneously on a solid substrate. The drop can be 
{}``divided'' into the following three regions: a {}``macroscopic'' bulk, a {}``mesoscopic'' film (a region that is within the range of surface forces), 
and a {}``microscopic'' precursor, the thickness of which amounts to several
molecular diameters. Note that figure 1 is schematic and the relative
sizes of these three regions are not up to scale. For brevity, we
will henceforth not use the adjectives {}``macroscopic'', {}``mesoscopic''
and {}``microscopic''; we will instead refer to the three regions
as the bulk droplet, the film and the precursor.

We note first that the edge of the precursor spreads well ahead of
the film and the bulk droplet: the radius of the precursor grows as
$\sqrt{t}$ \cite{pgc,ya,yab,yac,yad,gary1,gary2} and the film plays
the role of a reservoir feeding the precursor; this picture is valid as long as the said reservoir is far from being exhausted. Thus the precursor is decoupled from the rest of the drop and its
only role in the process can be seen as lubricating the substrate
for spreading of the film and the bulk drop.

A fundamental assumption of this approach is that the length scales
of the bulk and the film are well separated, i.e., that the bulk of
the drop is much wider and taller than the film. Thus the bulk can
be adequately approximated by a spherical cap with the base radius
$R(t)$, contact angle $\theta(t)$, and nearly constant
volume $V$ ($V=\frac{\pi}{4}R^{3}\theta$ for sufficiently small
$\theta$), i.e., the bulk is in equilibrium at constant volume $V$
and an instantaneous base radius $R(t)$.

Then, we can define an instantaneous free energy $2\pi R(t)F(t)$
along with an instantaneous rate of energy dissipation $2\pi R(t)W(t)$,
and propose that the evolution of these two quantities obeys the standard
relationship of the mechanics of dissipative systems: \begin{equation}
\left(W_{\mathrm{macro}}+W_{\mathrm{meso}}+W_{\mathrm{micro}}\right)R=-U\frac{\mathrm{d}\left(R\, F\right)}{\mathrm{d}R},\label{eq:QuasistationaryEnergyBalance}\end{equation}
where $U=\mathrm{d}R/\mathrm{d}t$ is the instantaneous velocity of
the apparent contact line. The right-hand-side (r.h.s.) of (\ref{eq:QuasistationaryEnergyBalance})
is the rate of change of the free energy of the system. It is equivalent
to the power of the driving force $\mathrm{d}F/\mathrm{d}R$ applied
to the moving contact line, and is balanced by the total dissipation
that occurs in the system, i.e., the left-hand-side (l.h.s.) of (\ref{eq:QuasistationaryEnergyBalance}).
The terms $W_{\mathrm{macro}}$, $W_{\mathrm{meso}}$ and $W_{\mathrm{micro}}$
are the dissipation rates in the bulk, film and precursor, respectively,
divided by the length $2\pi R$ of the apparent contact line.
Keep in mind that $F(t)$, $W(t)$ and the driving force $\mathrm{d}F/\mathrm{d}R$
are also, by definition, reduced by the length $2\pi R(t)$.

Next we specify the r.h.s. of (\ref{eq:QuasistationaryEnergyBalance})
\cite{degennes85}: \begin{equation}
-\frac{1}{R}\frac{\mathrm{d}\left(R\, F\right)}{\mathrm{d}R}\simeq S+\sigma\left(1-\cos\theta\right).\label{eq:dFdRcapillary}\end{equation}
Equation (\ref{eq:dFdRcapillary}) takes into account the surface
energies of the three interfaces meeting at the macroscopic contact
line and determines their variation with respect to $R(t)$ at constant
spherical cap volume. Note that the result is equivalent to a straightforward
application of the Young law; in fact, the r.h.s. of (\ref{eq:dFdRcapillary})
is typically referred to as an {}``unbalanced Young force''.

As for the l.h.s. of (\ref{eq:dFdRcapillary}), we can formally decompose
the dissipation according to the regions outlined in figure \ref{fig:DropCrossSection}a,
as follows: $W_{\mathrm{macro}}$ corresponds to hydrodynamic dissipation
in the bulk drop, where viscous flows are driven by the capillary
pressure; $W_{\mathrm{meso}}$ corresponds to hydrodynamic dissipation
in the film, where viscous flows are driven by the disjoining pressure;
$W_{\mathrm{micro}}$ corresponds to friction at the microscopic scale,
both at the edge of the film and in the molecular precursor.

The dissipation in the bulk drop is well-approximated by that in a
wedge, and is of the form \begin{equation}
W_{\mathrm{macro}}\simeq3\,\eta\, U^{2}\, g(\theta)\,\ln\left|\frac{x_{\mathrm{max}}}{x_{\mathrm{min}}}\right|,\label{eq:TSw}\end{equation}
where $\eta$ is the viscosity, $x_{\mathrm{max}}$ and $x_{\mathrm{min}}$
are effective cutoff lengths for the integration over the droplet
height and $g(\theta)$ is a known function of the instantaneous contact
angle. A salient feature is that the leading asymptotic behavior of
$g(\theta)$ when $\theta\to0$ is $g(\theta)\simeq1/\theta$, which
means that $W_{\mathrm{macro}}$ exhibits an unbounded growth as $\theta\to0$.
In the following we shall use the notation $\kappa=3\,\ln\left|\frac{x_{\mathrm{max}}}{x_{\mathrm{min}}}\right|$:
this is a slow-varying, empirical quantity, which varies only slightly
as the droplet spreads and introduces minor, logarithmic corrections
to the power laws; experimental data suggest that a good choice is
a nearly-constant $\kappa\approx120$ \cite{caz}.

We now come to the dissipation in the film. A striking result of Hervet
and de Gennes \cite{hervet} is that the complete wetting regime is
characterized by \begin{equation}
W_{\mathrm{meso}}\simeq S\, U,\label{eq:TSf}\end{equation}
which means that the dissipation within the film compensates exactly
the first term on the r.h.s. of (\ref{eq:dFdRcapillary}), rendering
the rate of spreading independent of $S$. This result was obtained
for non-retarded van der Waals substrate forces, but can be generalized.

Finally, the form of the dissipation term $W_{\mathrm{micro}}$ was
discussed by Blake and Haynes \cite{blake,blake2}: it was found that
$W_{\mathrm{micro}}\sim\zeta U^{2}$ at leading order in $U$, where
$\zeta$ is a constant friction coefficient. Note that in the case
of complete wetting $\zeta$ is dependent on the thickness of the
precursor.

Thus, provided that $S$ is consumed entirely in the film, the dynamical
behavior results from a competition between the two remaining dissipation
channels, $W_{\mathrm{macro}}$ and $W_{\mathrm{micro}}$. As $\theta\rightarrow0$,
$W_{\mathrm{micro}}$ is $\theta$-independent, whereas $W_{\mathrm{macro}}\sim1/\theta$
and thus clearly dominates at long spreading times. Dissipation at
the microscopic contact line may dominate (e.g., for low viscosity
liquids) at intermediate times, but ultimately hydrodynamic dissipation
in the core drop will take over \cite{degennes85,we}.

Consequently, neglecting $W_{\mathrm{micro}}$ as compared to $W_{\mathrm{macro}}$,
one finds in the small-$\theta$ limit that (\ref{eq:QuasistationaryEnergyBalance})
adopts the following form:

\begin{equation}
\theta^{3}\approx2\,\kappa\,\mathrm{Ca},\label{tanner1}\end{equation}
which is a fundamental relation between the velocity of the moving
contact line and the instantaneous value of the contact angle ($\mathrm{Ca=\eta U/\sigma}$
is known as the capillary number). It was derived analytically by
Voinov \cite{voinov} and by Tanner \cite{tanner} using a different
approach (within the lubrication approximation).

Now taking into account that the volume $V\approx\frac{\pi}{4}R^{3}\theta$
of the bulk drop remains nearly constant during spreading, we obtain
\begin{equation}
\dot{R}=\frac{64}{\pi^{3}}V^{3}\left(\frac{\kappa\eta}{2\sigma}\right)^{-1}R^{-9},\label{eq:tanner}\end{equation}
 from which the Tanner laws $R\sim t^{1/10}$ and $\theta\sim t^{-3/10}$
ensue trivially. The behavior described by (\ref{tanner1}) and (\ref{eq:tanner})
has been observed experimentally in \cite{tanner,aus,caz}.

\section{Energy balance approach revisited}

Equations (\ref{eq:dFdRcapillary}), (\ref{tanner1}) and (\ref{eq:tanner}),
under the assumption of well separated length scales of the bulk and
film, predict an unbounded growth of $R$. This is an ideal spreading
behaviour, through which the droplet virtually thins down to a molecular
film. As we have already remarked, this is not always the case, as
the spreading may terminate with the appearance of equilibrium pancake-like
structures \cite{degennes85,ruck,joanny1984}. In the latter situation,
the Tanner law clearly describes an intermediate stage, and the transition
to a pancake must be described by some kind of crossover in terms
of $R(t)$: our intuition is that $R$ will tend to a finite value
although it is not clear whether the definition of $R$ will remain
consistent with figure 1a.

An opposite trend was revealed by recent experimental studies focusing
on the spontaneous spreading of nematic liquid crystals (cyanobiphenyl
5CB) on hydrophilic \cite{cazb} or hydrophobic \cite{cazc} substrates:
after a transient Tanner stage, an {}``acceleration'' of the spreading
process has been observed. The data suggest that the base radius
$R$, as inferred from the inflection point of the thickness profile,
grows algebraically but with an exponent which is substantially larger
than $\alpha_{\mathrm{T}}=0.1$. In \cite{cazb} it was shown that
the Tanner law crosses over to $R\sim t^{\alpha}$ with $\alpha\approx0.2$.
Later it was realized (see figure 6 in \cite{cazc}) that the Tanner
relation in (\ref{tanner1}) does not hold for late stages of spreading:
for small $\theta$ and $\mathrm{Ca}$ the best fit to the experimental
data follows $\theta\sim\mathrm{Ca}^{0.7}$ rather than that predicted
by (\ref{tanner1}). The latter relation, together with the volume
conservation condition $R^{3}\theta\sim V$ yields $R\sim t^{\alpha}$
with $\alpha\approx0.19$. A more thorough analysis of the behavior
depicted in figure 5 in \cite{cazc} suggests that actually the reported
law $R\sim t^{1/5}$ is only a part of a crossover from the Tanner
stage to an even faster growth law. Experimental data in \cite{cazc}
span time scales ranging from a second to two hours, and at the end
of the experiment the trend is clearly rather $R\sim t^{1/3}$ than
$R\sim t^{1/5}$, and possibly still accelerating. Consequently, although
there is no conclusive evidence on the precise value of the exponent
$\alpha$ characterizing the accelerated spreading regime, it is clear
that $\alpha$ is significantly larger than the Tanner exponent and
thus the physical mechanism responsible for the late, post-Tanner
stages of spreading might be different from the one described in section
2.

\subsection{First guess: shear thinning}

We notice first that the accelerated spreading was observed for nematic
liquid crystals. Nematic crystals are known to have a non-Newtonian,
shear-thinning rheology (see, e.g., \cite{shear,nakano2003}). Shear
thinning affects the flow pattern, which necessarily modifies the
spreading dynamics. Thus our first idea is to revisit the l.h.s. of
(\ref{eq:QuasistationaryEnergyBalance}) and, more specifically, the
term $W_{\mathrm{macro}}$ in (\ref{eq:TSw}). The expression (\ref{eq:TSf})
for $W_{\mathrm{meso}}$ is also queried in the case of shear thinning.

A detailed analysis of the contact line dynamics within the framework
of the thin film model shows that the characteristic shear rates in
the capillary wedge and in the film \emph{decrease} as the contact line 
velocity decreases \cite{carre}. In consequence, for a
spontaneously spreading droplet of a non-Newtonian, shear-thinning
fluid the effective viscosity will \emph{increase} with time resulting
in a spreading law of the form $R\sim t^{\alpha}$ with $\alpha<1/10$.
Numerical simulations carried out in \cite{starov} confirm that $\alpha<1/10$
for shear-thinning fluids and that $\alpha>1/10$ for shear-thickening
fluids. Hence the dominant effect from shear thinning is that the
spontaneous spreading of a non-Newtonian fluid is generally \emph{slower}
than predicted by the Tanner law and can not explain the experimentally
observed acceleration of the spreading process.

\subsection{Second guess: line tension}

We now turn our attention to the r.h.s. of (\ref{eq:QuasistationaryEnergyBalance})
and notice that the unbalanced Young force - which is also the r.h.s.
of (\ref{eq:dFdRcapillary}) - is, in fact, a mere approximation of
the actual driving force of spreading. In general, the total free
energy of the liquid/solid system can be decomposed into bulk, surface,
line, and point contributions (see, e.g., \cite{getta,schimmele07}).
Thus, the driving force (\ref{eq:dFdRcapillary}), as a derivative
of the total free energy, should also contain all these contributions. 
This picture, of course, is meaningful only in the case of macroscopic drops, 
for which both the surface tension and the line tension are well defined.

As a matter of fact, the spherical cap adequately describes the profile
of the bulk drop, but the profile of the mesoscopic film deviates
from it, such that the quasistationary free energy $2\pi R\, F$ must
include a correction term, which is accumulated in the vicinity of
the apparent contact line and can be seen as a \emph{line} energy
$\tau$ multiplied by the apparent perimeter $2\pi R$ \cite{widom}.
The expression (\ref{eq:dFdRcapillary}) takes into account the surface
energies of the three macroscopic interfaces meeting at the apparent
contact line, but does not include the line tension contribution.

The idea that the line tension may have an appreciable impact on the
global behavior is not new. As an excess quantity, $\tau$ can be
positive or negative, as noticed already by Gibbs \cite{gibbs}. Negative
line tension, for example, can significantly reduce the work required
to create a nucleus (100 Angstrom in diameter) of a new phase on solid
or liquid substrates \cite{sheludko}. Conversely, positive line tension
can explain the stability of Newtonian black films towards rupture
\cite{kashiev}. For liquid droplets of nanometer size, negative (resp.
positive) line tension can promote spreading (resp. dewetting) even
if the macroscopic spreading parameter $S$ is negative (resp. positive)
\cite{widom}. A review of different phenomena caused by line tension
effects and some conceptual aspects of line tension can be found in
\cite{getta,schimmele07}.

Evaluation of the contribution due to the line tension $\tau$ involves
many delicate issues (e.g., a proper definition of the effective interface
potential used in the model, or a proper convention when choosing
the Gibbs dividing interface) and, in general, is a more complex problem
than the calculation of the surface tension - essentially because
more phases meet at the contact line than at an interface (see,
e.g., \cite{getta,schimmele07} for a more thorough discussion). The
problem is already difficult in equilibrium situations (e.g., partial
wetting, with $S<0$), and clearly becomes even more complex when
one considers spontaneous spreading, since here one has to account for
the temporal evolution of the droplet thickness profiles.

The consideration of these subtle points is beyond the scope of the
present approach. For our purposes it will be sufficient to resort
to a recently proposed phenomenological generalization of (\ref{eq:dFdRcapillary})
in terms of non-equilibrium thermodynamics: as shown in \cite{fan},
the force applied to the apparent contact line of a droplet can be
written down as \begin{equation}
f_{\tau}=S+\sigma\left(1-\cos\theta\right)-\frac{\tau}{R},\label{force1}\end{equation}
which differs from the expression in (\ref{eq:dFdRcapillary}) by
an additional term accounting for the contribution of the contact
line tension $\tau$ to the driving force of spreading. The definition
of $f_{\tau}$ as a generalized Young force is valid both in complete
and partial wetting, and is consistent with the so-called modified
Young equation $f_{\tau}=0$, obtained at equilibrium by an appropriate
generalization of Gibbs classical theory of capillarity \cite{widom,boruvka}.

At this point we must stress that the expression (\ref{force1}) for $f_{\tau}$
is formally valid only if $\tau$ is constant. If we assume a power-law
behaviour for $\tau\sim R^{\beta}$, then the modified Young force
becomes $S+\sigma\left(1-\cos\theta\right)-\frac{\tau}{R}\left(1+\beta\right)$.
Here we argue that it is not likely for $\tau$ to vanish at long
$t$ and large $R$, and thus $\beta\geq0$. Thereby (\ref{force1})
essentially holds for $\tau\sim R^{\beta}$, up to a numerical factor
$(1+\beta)>0$ applied to the line tension term, i.e., the last term
on the r.h.s. of (\ref{force1}). This consideration has little impact
on the following qualitative argument, but will be relevant to quantitative
implications.

\subsection{Line tension effects on spreading}

Suppose now that (\ref{eq:QuasistationaryEnergyBalance}) holds; that
the dissipation $W_{\mathrm{meso}}$ in the film obeys (\ref{eq:TSf});
that the dissipation $W_{\mathrm{macro}}$ is given by (\ref{eq:TSw});
but that the driving force of spreading is now determined by (\ref{force1}),
i.e., that the line tension contribution is taken into account . Then
(\ref{tanner1}) is replaced with the following:\begin{equation}
\kappa\frac{\eta}{\theta}\dot{R}=\frac{1}{2}\sigma\theta^{2}-\frac{\tau}{R}.\label{eq:QuasistationaryEnergyBalance3}\end{equation}

Since $\tau$ is typically very small, one naturally finds that the
surface tension contribution will dominate at small and intermediate
times, again giving rise to the Tanner law $R\sim t^{1/10}$. On the
other hand, as $R$ grows to sufficiently high values, the line tension
contribution will inevitably take over and become a dominant driving
force, provided that $\tau/R$ decays slower than the capillary term
$\sigma\theta^{2}$.

In the latter regime, the behavior is crucially dependent on the sign
of $\tau$:
\begin{itemize}
\item If $\tau$ is positive and tends to a constant value, which is physically
plausible, then (\ref{eq:QuasistationaryEnergyBalance3}) predicts
that spreading will terminate at a finite value of $R$, for which
the first and the second terms on the r.h.s. of (\ref{force1}) become
equal to each other. One may interpret this as an indication of the
formation of a pancake. However, during the actual transition to a
pancake, a drop would not retain the shape of a capillary cap (see
figure 1b), which somewhat challenges this prediction.
\item If $\tau$ is negative and the second term in (\ref{eq:QuasistationaryEnergyBalance3})
dominates, we find the following post-Tanner behavior: \begin{equation}
R\sim\left(-\int^{t}\,\tau dt\right)^{1/5}.\label{gen}\end{equation}
This spreading law is qualitatively different from $R\sim t^{1/10}$.
In the framework outlined in section 2, the driving force of spreading
is associated with the surface tension. By contrast, during the late stages
of spreading, the droplet becomes flatter and can be viewed as {}``quasi
two-dimensional''. It is then not surprising that the line tension
$\tau$ should govern the spreading process, provided that $\left|\tau\right|\gg\sigma R\theta^{2}$.
\end{itemize}
It is tempting to obtain coarse quantitative results from (\ref{gen})
and from the condition $\left|\tau\right|\gg\sigma R\theta^{2}$.
In particular, if we assume that $\tau$ is a negative constant, then
(\ref{gen}) predicts $R\sim t^{1/5}$, which agrees with previously
reported experimental results \cite{cazb,cazc}. Looking at figure
5 in \cite{cazc} we can estimate the value of $\tau$ from the characteristic
base radius at the apparent crossover between the Tanner stage and the
accelerated spreading regime: this yields $\tau\approx-10^{-9}$ N.
This value is an order of magnitude higher than previously reported
values of the line tension in the partial wetting situations, but it must 
be noted that many experimental measurements of $\tau$ have been performed 
for simple liquids; in the case of nematic liquid crystals an elastic
contribution to the effective interface potential (a consequence of the 
anchoring properties) may yield substantially higher values of $\tau$.

However, upon a closer examination of the case of negative $\tau$, 
there is no good reason to expect that $\tau$ should approach a constant
value. One rather expects that the film region will become progressively
more pronounced and the drop profile will significantly deviate from
a spherical cap-like shape; hence $\tau$, a functional of local droplet
thickness, will grow as a function of time (in terms of its absolute
value). Consequently, at late spreading stages, one may expect a growth
of $R(t)$ that is faster than $R\sim t^{1/5}$.

\section{Conclusions}

We have presented in this paper both the classical energy balance
approach - as developed by de Gennes - and a revised version of it,
which incorporates a line tension contribution to the driving force
of spreading. The revisited framework was motivated by apparently
contradictory trends at long spreading times for macroscopic droplets in complete
wetting. By taking line tension into account, we have complemented
the classical framework with the following twofold interpretation:
\begin{itemize}
\item A positive line tension - essentially a {}``collar'' around a spreading
droplet - stops spreading and is responsible for the formation of
mesoscopic pancakes.
\item A negative line tension - which tends to lengthen the apparent contact
line - governs the late stages of spreading, resulting in the acceleration
of this process.
\end{itemize}
We must now voice a few words of caution concerning our approach,
which seems intuitive but may have several shortcomings due to its
simplicity.

First we admit that the formal definition of line tension $\tau$
in partial wetting - a functional of an equilibrium profile, and an
integral of the effective interface potential - cannot be easily generalized
to complete wetting and thus remained undetermined within our analysis.
The case of constant, positive $\tau$ is plausible, but negative
$\tau$ is more likely to grow as a function of time and base radius,
a growth which we are unable to specify. We merely expect that, as
in the partial wetting case, $\tau$ is a certain functional of the
mesoscopic thickness profile, rather than an independent, arbitrary
quantity (see, e.g., \cite{EPL}).

We have already stated that in the case of positive $\tau$ our prediction
of the emergence of pancakes is indicative at best: indeed, in the
process of reaching the stationary shape of a pancake, a droplet gradually
deviates from the spherical cap shape assumed by sections 2 and 3;
this deviation entails corrections that our approach does not account
for. A similar word of caution exists for negative $\tau$ and arises
from a thorough comparison with experiment. Indeed, a remarkable feature
of nematic 5CB droplets observed in \cite{cazb,cazc} is that the
reported accelerating phase is accompanied by the development of a
large {}``foot'' (essentially the drop adopts a bell shape similar
to figure 1a, without exaggeration). This large foot is a warning
against the applicability of energy balance as developed in sections
2 and 3: the key assumption of a spherical cap of constant volume
$V_{\mathrm{cap}}=\frac{\pi}{4}R^{3}\theta$ is less and less valid
as the mesoscopic film drains liquid from the macroscopic droplet.
In other terms, the separation of macroscopic and mesoscopic length
scales (both vertical and lateral), in the experimental layout that
we are trying to describe, may be more precarious than what was assumed
in our analytical framework.

In the light of these shortcomings, our agenda is to develop a more
robust approach describing late stages of droplet spreading, based
on the seminal approach by Tanner describing the time evolution of
the thickness profile of whole droplets within the thin film approximation
\cite{voinov,tanner}. In the latter framework, scale separation is
not an issue, and the fundamental result is that droplets gradually
turn into diffusive films in the sense of Derjaguin \cite{derjaguin}.
Results have already been obtained for the specific case of nematic
droplets, and we shall present them in a companion paper
\cite{companion}.

We are also looking forward to developing the notion of a dynamic line
tension in spreading processes, through a detailed study of the hydrodynamic
wedge as in \cite{degennes85,eggers}. It is important to note that
nematic 5CB droplets exhibit both of the non-trivial types of post-Tanner
behaviour detailed in this paper: $R(t)$ crosses over from $R\sim t^{0.1}$
to faster power laws (at this point the drop looks like figure 1a),
but eventually the spreading terminates with a mesoscopic pancake
(figure 1b). In terms of $\tau$, this means that the physically relevant
line tension is negative during the accelerating phase, but later
becomes positive. As figure 1 suggests, the geometrical properties
of the droplet are quite different in both regimes, and it is not
clear whether a consistent definition of $\tau$ can be established
for such liquids as nematic liquid crystals.

\ack{The authors gratefully acknowledge helpful discussions with
H. Tanaka}

\end{document}